\documentclass[lettersize,journal]{IEEEtran}
\usepackage{amsmath,amsfonts}
\usepackage{bm}
\usepackage{float}
\usepackage{graphicx}
\usepackage{multirow,enumitem}
\usepackage[mathscr]{eucal}
\usepackage{hyperref}
\usepackage[table,xcdraw]{xcolor}
\usepackage{footmisc}
 \usepackage{booktabs}
 \usepackage{caption}
\usepackage{subcaption}
\begin{document}

\title{HCAM - Hierarchical Cross Attention Model for Multi-modal Emotion Recognition}

\author{Soumya Dutta,~\IEEEmembership{Student Member,~IEEE,} and 
        Sriram Ganapathy,~\IEEEmembership{Senior Member,~IEEE}
        % <-this % stops a space
\IEEEcompsocitemizethanks{\IEEEcompsocthanksitem 
S. Dutta and S. Ganapathy are with the learning and extraction and acoustic patterns (LEAP) laboratory, Electrical Engineering, Indian Institute of Science, Bangalore, India, 560012. This work was performed with grants received as part of the prime ministers research fellowship (PMRF), Ministry of Education, India. 
\protect\\
E-mail: \{soumyadutta, sriramg\}\@ iisc.ac.in 
%\IEEEcompsocthanksitem J. Doe and J. Doe are with Anonymous University.
}% <-this % stops an unwanted space
\thanks{Manuscript received March xx, 2023.}}


\maketitle

\begin{abstract}
Emotion recognition in conversations is challenging due to the multi-modal nature of the emotion expression. 
We propose a hierarchical cross-attention model (HCAM) approach to multi-modal emotion recognition using a combination of recurrent and co-attention neural network models. The input to the model consists of two modalities, i) audio data, processed through a learnable wav2vec approach and, ii) text data represented using a bidirectional encoder representations from transformers (BERT) model. The audio and text representations are processed using a set of bi-directional recurrent neural network layers with self-attention that converts each utterance in a given conversation to a fixed dimensional embedding. In order to incorporate contextual knowledge and the information across the two modalities,  the audio and text embeddings are combined using a co-attention layer that attempts to weigh the utterance level embeddings relevant to the task of emotion recognition. 
The neural network parameters in the audio layers, text layers as well as the multi-modal co-attention layers, are hierarchically trained for the emotion classification task.
We perform experiments on three established datasets namely, IEMOCAP, MELD and CMU-MOSI, where we illustrate that the proposed model improves significantly over other benchmarks and helps achieve state-of-art results on all these datasets.  
\end{abstract}


\section{Introduction}
\IEEEPARstart{H}{uman} emotions, expressed in a complex multi-modal manner, play a central role in human interactions and inter-person  communications. 
As automated systems play a ubiquitous role in day-to-day lives, the understanding of the human emotions becomes a crucial step in the design of these systems.  
The machines capable of performing emotion recognition can enable the development of personalized human-computer interfaces like conversational agents \cite{pantic2005affective}, social media analytics~\cite{gaind2019emotion}, customer call centres \cite{li2019acoustic} and  mental health monitoring systems \cite{ghosh2019emokey}. 
The key task is termed as emotion recognition in conversation (ERC).
%, differs from a sentiment analysis task, where a single positive or negative sentiment is attached to a group of utterances \cite{kratzwald2018deep,colnerivc2018emotion}. 

Emotion recognition in conversational data entails a number of challenges, namely multiple (overlapping) speakers, short-term and long-term dependencies~\cite{poria2019emotion}, short duration of turn events, background noise in audio etc. Further, emotion recognition task is inherently multi-modal, where the information is expressed in a variety of ways such as facial expressions~\cite{tarnowski2017emotion}, speech~\cite{scherer2003vocal}, gestures~\cite{navarretta2012individuality}, physiological signals~\cite{knapp2011physiological} or through a combination of these. The different modalities contain varying degrees of information relating to the underlying emotion and hence, designing a joint multi-modal approach for  emotion recognition is usually considered~\cite{poria2017review}. 

The previous works have explored the use of  audio with visual signals ~\cite{kim2017isla,zhou2021information} and text with audio signals~\cite{yoon2018multimodal}, ~\cite{pepino2020fusion}. Further, there have been attempts to use all the three modalities namely visual, audio and text~\cite{mittal2020m3er}. While the ability to perceive emotions in a multi-modal way is required, it is also necessary to recognize the emotions from each modality in a robust manner, for scenarios where the data from some of the modalities is unavailable or noisy.
%This may become necessary in the scenarios where one may not have access to all the modalities for emotion recognition.

%In this paper, we explore the task of joint emotion recognition from audio and text. The fusion of the information streams from the two modalities is done by means of the widely popular co-attention mechanism.
% We extend our previous work on multi-modal transformer models \cite{dutta2022multimodal} using a novel model architecture  that facilities the fusion of the two modalities based on the principles of co-attention. The co-attention is a mechanism that jointly models  auditory attention and textual
% attention. Unlike previous works, which only focus on simple concatenation of information streams, the proposed approach, with a natural symmetry between the audio based and text based embeddings,  guides the model to attend to the multi-modal contextual information for emotion recognition. 

 In this paper, we propose a novel hierarchical cross attention model (HCAM) for the problem of multi-modal emotion recognition. Our modeling framework, involving large representation learning networks from speech and text, uses a hierarchical training process. The training paradigm consists of three distinct stages, where the first stage trains uni-modal predictors in a context-agnostic fashion. The contextual information, being an essential component in conversational emotion recognition, is added subsequently in the second stage. The final stage involves the fusion of the acoustic and textual streams using a co-attention module.
  The audio features are extracted using a wav2vec model \cite{baevski2020wav2vec}  while textual representations are derived using RoBERTa \cite{liu2019roberta}  model. The co-attention is a mechanism that jointly models  auditory and textual information. With a natural symmetry between the audio based and text based embeddings, this fusion technique guides the model to attend to the multi-modal contextual information for emotion recognition.    
% %We also propose a supervised contrastive loss along with the FoCal loss \cite{lin2017focal} for training the system.
%Fusion of multiple modalities are primarily of three types - early, late and hybrid. In early fusion, the features extracted from the individual modalities are combined (often concatenated) and fed to a module capable of this joint modeling. Late fusion involves training the individual modalities and then combining them at the decision level. Finally, the hybrid approach combines these two for the final decision. 
 In order to train this model in a hierarchical fashion, we explore the usage of the supervised contrastive loss~\cite{khosla2020supervised}.

The experiments are performed on three established datasets, namely IEMOCAP~\cite{busso2008iemocap}, MELD~\cite{poria2019meld} and CMU-MOSI~\cite{zadeh2016mosi}. Although the  type of emotions, duration of the utterances, length of the conversations, noise or distortion in the audio and style of the conversations are different across the datasets, the same HCAM  architecture is proposed.
% % We note that audio is the more commonly available signal to a personal agent. It is with this motivation that we train our models not only with the available ground truth transcripts but also with the transcripts generated from an Automatic Speech Recognition (ASR) system and use it to enhance the emotion recognition capability from the audio signal alone. We train separate uni-modal models for audio, ASR and text. For each modality, we further break the problem of ERC into two parts - we first treat each utterance separately irrespective of the conversation it is a part of, following which, we utilize the context from the conversation for emotion recognition.  We use the co-attention network for combining audio and ASR features to achieve a audio signal only accuracy. The combined ASR and audio features are then combined with the text uni-modal model embeddings by means of a combination of self-attention and co-attention networks to predict the emotions for each utterance in the conversation. This model architecture is found to be more accurate in predicting the most frequent classes in ERC datasets. As ERC datasets are in general imbalanced, we train our model by the focal loss to improve the overall performance of our model. This loss, aimed at classifying each sequence of utterance in a conversation correctly, however, ignores the hypothesis that representations of utterances belonging to the same emotional class from different conversations must be similar and therefore we train our models with the focal loss along with the supervised contrastive loss. 

The key contributions from this work are:
\begin{itemize}
    \item We propose a hierarchical approach for the multi-modal emotion recognition, where the information is first processed at the utterance level in each of the modalities, followed by inter-utterance conversation  modeling and subsequently, multi-modal processing with cross-attention.  We experimentally establish that the hierarchical modeling is important for improving the emotion recognition performance. 
    \item As the model learns different aspects of the conversations in every stage, we propose to combine the model predictions from the previously trained stages and the current stage during test time to further improve the performance
    \item We use the supervised contrastive loss in addition to cross-entropy loss in model training. The supervised contrastive loss allows the model to focus more on the  hard training examples. 
    \item We test the proposed model on three  benchmark datasets and achieve state-of-the-art results for all these datasets. Further, we also experiment with replacing the ground-truth text with ASR transcripts during test time in order to test the robustness of our model.
    %The experiments illustrate that the proposed HCAM approach improves significantly over other benchmarks and achieves state-of-art results on all the three datasets.
    %
\end{itemize}\par

%The rest of the paper is organized as follows. Section \ref{relatedworks} discusses some related works on SER and text sentiment analysis followed by works on ERC. Section \ref{background} details the background behind this work while section \ref{proposal} discusses the details of the proposed architecture. We talk about the different datasets and experimental settings in which the proposed models are tested along with the results obtained in section \ref{experiments} and the key results are subsequently analyzed in section \ref{discuss}. Finally we summarize the work in section \ref{conclude}.
\begin{figure*}
    \centering
    \includegraphics[width=\textwidth,trim={0cm 0cm 0cm 0cm},clip]{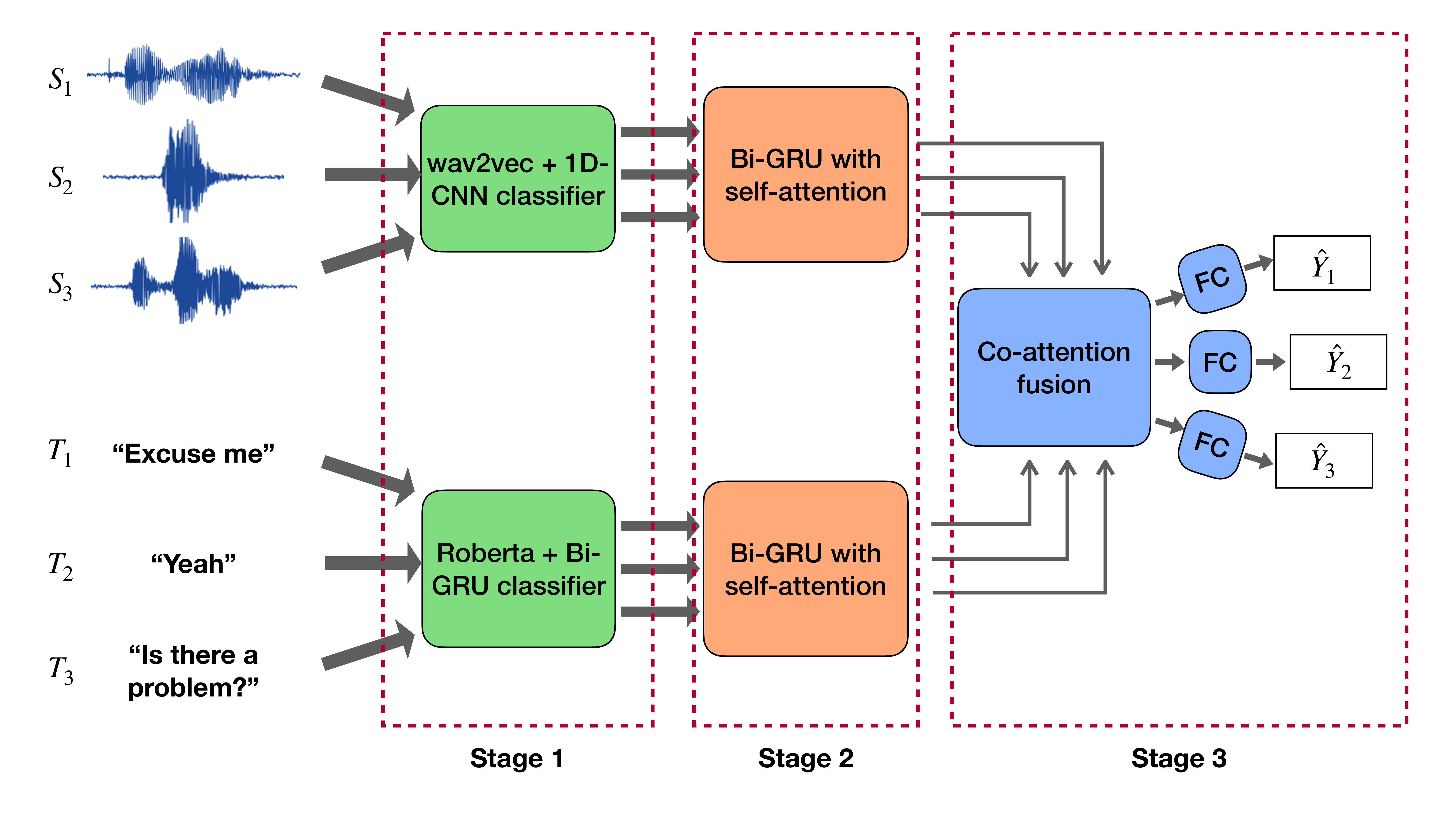}
    \vspace{-0.1in}
    \caption{Block diagram of the proposed model. Here, $S_1$, $S_2$ and $S_3$ refer to the speech utterances in a conversation. Similarly, the $T_1$, $T_2$ and $T_3$ refer to the text transcripts for the corresponding speech signals. $\hat{Y_1}$, $\hat{Y_2}$ and $\hat{Y_3}$ refer to the predicted emotion labels for the three utterances. The three stages of training are also marked in the diagram.}% The wav2vec2.0 and CNN block for the audio signals along with the RoBERTa-Bi-GRU networks for ASR and text constitute the first stage of training while the three Contextual GRU blocks are trained in the second stage. The co-attention blocks are trained for combining audio and ASR information. Finally, text is combined with the fused audio-ASR representation by means of another co-attention block. Self-attention blocks are also used on the fused audio-ASR representation and the text representations. These are concatenated and passed through position-wise fully connected layers for classification.}
    \label{fig:entire model}
    \vspace{-0.1in}

\end{figure*}

\section{Related Works}\label{relatedworks}
\subsection{Audio feature extraction}
The initial approaches for emotion recognition used features that were mostly knowledge driven like amplitude, pitch and spectral profile, as proposed by Sauter et. al. ~\cite{sauter2010perceptual}. Luengo et. al.~\cite{luengo2005automatic} used prosodic parameters for SER, while segment level prosodic features were used by Koolagudi et. al.~\cite{koolagudi2011speech}. The mel-frequency cepstral coefficients (MFCCs) were highlighted to provide the best emotion classification performance by Eyben et. al~\cite{eyben2013affect}. Recently, the statistical descriptors of a number of knowledge driven features were found to be significantly better for the SER task by Schuller et. al, as part of the Interspeech para-linguistics challenge~\cite{schuller2013interspeech}. These features were further refined~\cite{eyben2015geneva} to create a minimalist set of parameters (Opensmile toolkit~\cite{eyben2010opensmile}). 

The progress in deep learning in the recent years have motivated researchers to develop audio feature extractors which are learnable in nature, like the SincNet features by Ravanelli et. al. \cite{ravanelli2018speaker}, learnable audio front-end (LEAF) by Zeghidour et. al. \cite{zeghidour2021leaf} and interpretable Gaussian filters by Agrawal et.al~\cite{agrawal2019modulation,agrawal2020interpretable}. Recently, unsupervised and self-supervised approaches have been proposed using a large corpus of unlabeled speech data~\cite{schneider2019wav2vec, baevski2020wav2vec}. 
%These models use contrastive predictive coding (CPC)  or  auto-regressive loss functions to learn the audio representations. 
Several recent works in emotion recognition have explored the use of these representations \cite{siriwardhana2020jointly, macary2021use, pepino2021emotion}.

\subsection{Text feature extraction}
Early approaches for text feature extraction used the bag-of-words approach  in conjunction with word relation features, as proposed by Xia et. al.~\cite{xia2010exploring}. The recent models, inspired by deep learning, use prediction tasks to learn text embeddings. One such attempt, termed the word2vec model by Mikolov et. al~\cite{mikolov2018advances}, is widely used for feature extraction in text sentiment analysis~\cite{majumder2018multimodal, sharma2017vector, liu2017sentiment}. With the development of recurrent and attention networks such as bidirectional long short term memory networks (B-LSTM), gated recurrent unit (GRU) and transformers~\cite{vaswani2017attention}, improved language models were developed like BERT~\cite{devlin2019bert} and RoBERTa~\cite{liu2019roberta}. The BERT embeddings have resulted in improvements for a variety of downstream tasks like sentiment analysis~\cite{lian2022smin}.

\subsection{Multi-modal fusion}
The fusion of multiple modalities has been shown to be effective in emotion recognition~\cite{poria2017context}. The early attempts using  a concatenation of the representations has been replaced with sophisticated techniques like multi-modal transformers~\cite{tsai2019multimodal, dutta2022multimodal}. A multi-view   sequential learning architecture was proposed by Zadeh et. al~\cite{zadeh2018memory}. This was further improved using a dynamic fusion graph~\cite{zadeh2018multimodal}. The attention based fusion approach is also pursued in various vision-language tasks~\cite{lu2019vilbert}, \cite{tan2019lxmert}.
%Mittal et. al~\cite{mittal2020m3er} proposed a multiplicative fusion strategy.
\subsection{Incorporating contextual information in ERC}
The presence of multiple speakers in a conversation may result in short-term and long-term dependencies. This presents a significant challenge in ERC, as highlighted by Poria et. al~\cite{poria2019emotion}. To address this, Poria et. al~\cite{poria2017context} used LSTM networks to capture inter-utterance context in the conversation. This was further refined by Majumder et. al~\cite{majumder2019dialoguernn} in their model, named DialogueRNN. The work by Ghosal et. al proposed the use of graph convolutional networks~\cite{ghosal2019dialoguegcn}. 
%Although, there have been a fairly large number of attempts to solve the problem of introducing context in ERC, it still remains a challenging task. Emotional shifts between one utterance to another is particularly difficult to model in these datasets.
\subsection{Contrast with prior works}
In contrast to the prior works, the proposed HCAM framework is novel in the following aspects. 
\begin{itemize}
    \item We propose a curriculum learning~\cite{bengio2009curriculum} based design of the modeling stages.  In this design, the easy task of recognizing the emotion states at an utterance level of a conversation is learned initially. The more complex task of inter-utterance contextual modeling is designed on top of the utterance level model with recurrent layers. Further, the neural attention based modeling layers enable the multi-modal integration. This hierarchical modeling framework is shown to efficiently learn the underlying emotion labels from speech and text representations.
    %\item Second, we propose a combination of self-attention and cross-attention modules in the multi-modal fusion. This combination enables the HCAM improve over the best performing individual modality on all the tasks considered.
    \item The supervised contrastive loss is explored for emotion recognition task. This loss  improves the learning capacity of the modeling by focusing on the harder training examples. 
    \item The ability of the model to classify emotions from a conversation varies in accordance with the modeling stage. We propose to ensemble the model predictions at different stages during inference.
    %\item Third, we evaluate the HCAM network on multiple speech-text emotion recognition datasets and establish new state-of-art results on these tasks.
\end{itemize} 

\section{Background}\label{background}
% In this section, we discuss the background  works on wav2vec~\cite{baevski2020wav2vec}, BERT~\cite{liu2019roberta} and attention schemes. These components form part of the proposed model, discussed in the next section. 
\subsection{wav2vec} 
The wav2vec is a representation learning framework based on principles of self-supervision \cite{schneider2019wav2vec}. In the wav2vec 2.0 model~\cite{baevski2020wav2vec}, the audio signal is windowed into short overlapping frames.   Each windowed segment is  passed through convolutional feature extractor layers, following which a quantization module allows encoding of the representations in a discrete space. These representations are  contextualized  by means of transformer encoder layers. The network is pre-trained with a self-supervised  contrastive loss. 

\subsection{RoBERTa}
In the recent years, one of the significant contributions towards creating a large scale language model, is by Devlin et.al~\cite{devlin2019bert}. This architecture, called bidirectional encoder representations from transformer (BERT), was trained with two tasks on a corpus of textual data, namely, predicting a word masked out in a sentence (called masked language modeling) and to predict whether two sentences semantically follow each other (referred to as next sentence prediction). Liu et. al~\cite{liu2019roberta} trained this architecture on a larger corpus of textual data after removing the next sentence prediction task. This pre-trained model, known as robust optimized BERT approach (RoBERTa), also optimizes several other hyper-parameters in the design of the model. 

\subsection{Self and cross attention}\label{sec:self_cross_attn} 
The self-attention mechanism, as proposed by Vaswani et. al~\cite{vaswani2017attention}, considers a sequence of length $N$, and of dimension of $d_k$. This sequence is converted into three matrices, $Q \in \mathcal{R}^{N \times d_k}$, $K \in \mathcal{R}^{N \times d_k}$ and $V\in \mathcal{R}^{N \times d_k}$. Self-attention is involved in computing the similarity between query representation (denoted by $Q$) with key representation (denoted by $K$). This similarity matrix, converted to a probability distribution by the softmax function, is then used to take a weighted sum of the value representations (denoted by $V$). Finally, the query matrix ($Q$) is added to this weighted sum followed by a layer normalization block.
    \begin{eqnarray}
    &Attention(Q, K, V) = softmax(\frac{QK^{T}}{\sqrt{d_k}})V\\\label{selfattn1}
    &Q = Q + Attention(Q, K, V)\\\label{selfattn2}
    &output = LayerNorm(Q)\label{selfattn3}
\end{eqnarray}
\par
% Attention networks have also been extensively used in problems where we require alignment between two different modalities. This was explored in the vision-language domain by Lu et.al~\cite{lu2019vilbert}. The use of cross-attention in \cite{lu2019vilbert} ensured similarity of representations across modalities to be considered. This novel architecture helped the authors achieve state-of-the-art results in a number of vision language tasks such as visual question answering~\cite{antol2015vqa}, caption based image retrieval~\cite{young2014image} and visual commonsense reasoning~\cite{zellers2019recognition}. 
The cross-attention network is similar to the self-attention module with a key difference. Here, the query matrix and the key/value matrices are constructed from representations of different modalities. If we consider the query matrix from audio modality (denoted as $Q_A$), and the key/value matrices from the text modality (denoted as $K_T$/$V_T$), the cross-attention network operations are, 
%The cross-attention block used in the model is shown in Fig. \ref{fig:attention}.
\begin{eqnarray}
&Atten(Q_A, K_{T}, V_{T}) = softmax(\frac{Q_{A}K_{T}^{T}}{\sqrt{d_k}})V_{T}\label{crossatt1}\\
&Q_A = Q_A + Atten(Q_A, K_{T}, V_{T})\label{crossatt2}\\
&Q_A = LayerNorm(Q_A).\label{crossatt3}
\end{eqnarray}

% \subsection{Focal Loss}
% The FoCal loss function aims to address classification problems on imbalanced classes~\cite{lin2017focal}.
% Let us consider the classification problem with $C$ classes, where the ground truth one-hot vector for a particular example is denoted by $\mathbf{y}=[y_1,y_2, \dots ,y_C]$. Further, assume that the example belongs to the class $j$.  
% %The cross-entropy (CE) loss for the example with ground truth class denoted by $j \in \{1\dots C\}$, with the prediction probability vector denoted by $\mathbf{\hat{y}}$ is defined as
% %\begin{equation}\label{ce}
% %    CE(\mathbf{y}, \mathbf{\hat{y}}) = -log(\hat{y_j}) - \sum_{k=1;k \neq j}^{C}log(1-\hat{y_k})
% %\end{equation}
% The FoCal loss is then given by,
% \begin{equation}\label{focal}
%     FL(\mathbf{y}, \mathbf{\hat{y}}) = -(1-\hat{y_j})^\gamma log(\hat{y_j}) %- \sum_{k=1;k \neq j}^{C}\hat{y_k}^\gamma log(1-\hat{y_k})
% \end{equation}
% where, $\mathbf{\hat{y}}$ denotes the model output. 
% %For a particular example, if the prediction probabilities for the different classes is given by $p$, $p_y$ is defined as $p$ for the ground 
% %In equation \ref{focal}, $p_y$ denotes the prediction probability of the ground truth class
% %the prediction probability of the ground-truth class is denoted by $p_y$. 
% As the model is expected to predict the frequent classes correctly with more confidence, Eq.(\ref{focal}) ensures that the loss corresponding to such examples are down-weighted (for $\gamma > 1$).
\section{Proposed Architecture}\label{proposal}

The block diagram of the proposed model is shown in Fig. \ref{fig:entire model}. As the entire model is a relatively large architecture, driven by the two modalities of text and audio, our model has three distinct stages which are trained hierarchically, where each stage uses the pre-trained model parameters of the previous stage without fine-tuning. 

%Individual feature extractors for audio and text/ASR are detailed in subsections \ref{audfeat} and \ref{textfeat} respectively. 
%The utterance-level embeddings obtained from this modeling stage are used in the next stage, where we introduce inter-utterance context by means of a bidirectional gated recurrent unit (Bi-GRU) architecture with self-attention mechanism (Sec.~\ref{context}). 
 All the stages are trained hierarchically using a weighted combination of cross-entropy loss and the supervised contrastive loss function (Sec.~\ref{loss}). 
The model implementation is made available publicly\footnote{\url{https://github.com/iiscleap/Multimodal_emotion_recognition_with_coattention}}. 
\begin{figure}[t!]
    \centering
    \includegraphics[width=0.5\textwidth,trim={8cm 3cm 15cm 3cm},clip]{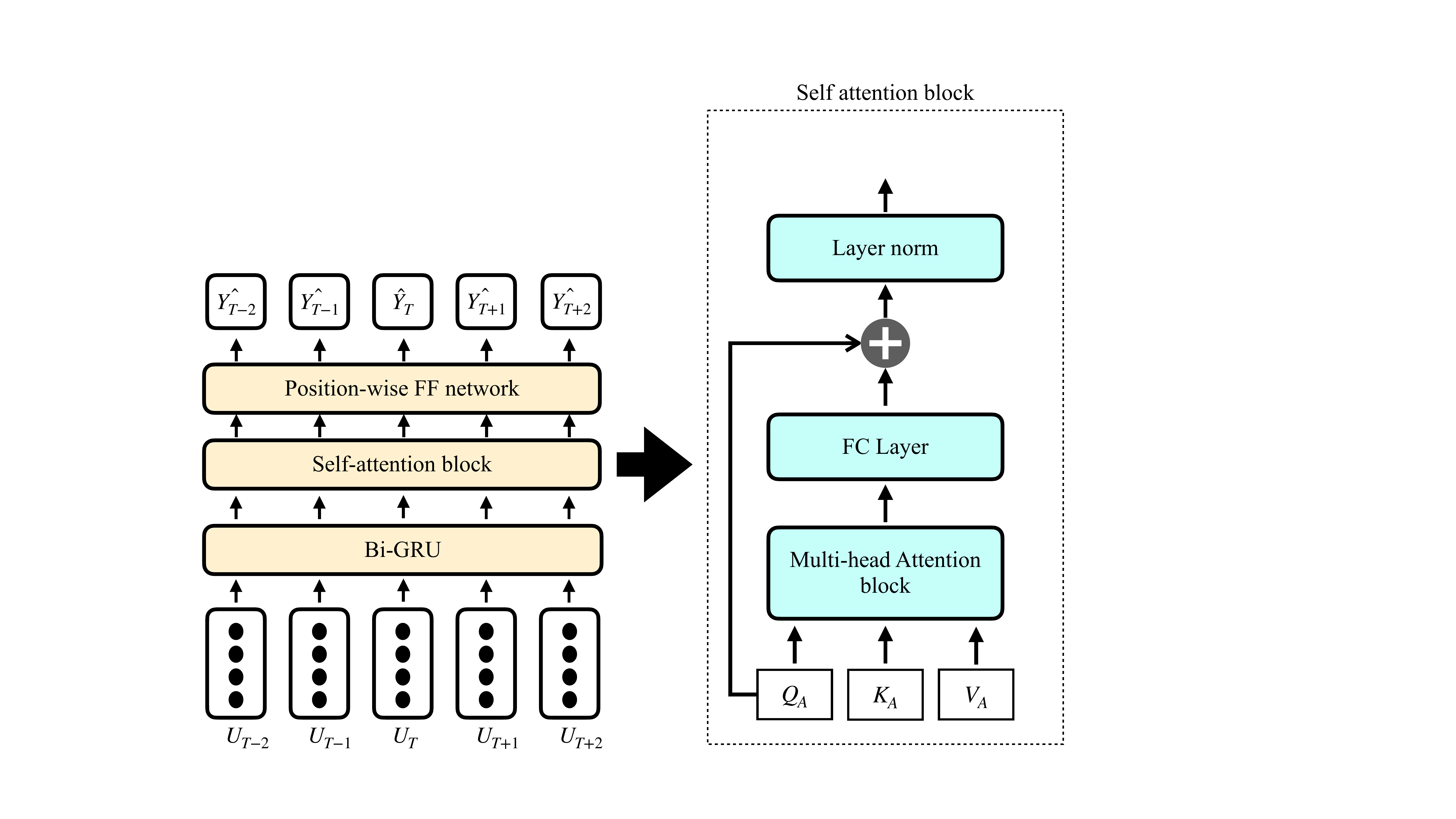}
    \caption{Block diagram of the contextual GRU with self-attention. Here, $U_{T}$, $U_{T \pm 1}$ and $U_{T \pm 2}$ refer to the uni-modal embeddings from stage I of the model for each utterance in the conversation.}
    \label{fig:context gru}
\end{figure}
\subsection{HCAM Stage I}\label{stage1}
In the first stage, we train utterance-level embedding extractors from audio and text. The models are trained to classify individual utterances without considering the inter-utterance conversational context. The models trained in this stage classify the individual utterances into the corresponding emotion classes based on cues present in either the audio signal or the text transcript. The pre-trained feature extractors are generally models with large computational requirements, and this constrains the number of utterances that can be processed in every iteration. The large number of utterances in a single conversation inhibits the fine-tuning of these feature extractors in previous works, as they process conversations as a whole. Our hierarchical modeling allows us to fine-tune the embeddings from the pre-trained feature extractors for improved emotion classification for each individual utterance in this training stage.
% (say, $B_u$). Previous works which handle a batch of conversations would need to process all the utterances in the batch. If the number of conversations being handled in every batch is $B_c$ and each conversation has number of utterances as $N_u$, the effective batch size for the pre-trained feature extractor becomes $B_c \times N_u$, which will be much larger than $B_u$ for most ERC datasets. We now provide a brief description of the audio and text feature extractors used in the proposed model.

\subsubsection{Audio embedding extractor}\label{audfeat}
The audio is input to the wav2vec2.0 large model~\cite{baevski2020wav2vec}, pre-trained on Libri-light~\cite{kahn2020libri}, CommonVoice~\cite{ardila2020common}, Switchboard~\cite{godfrey1992switchboard} and Fisher~\cite{cieri2004fisher} datasets. The model is further fine-tuned on $300$ hours of noisy telephone conversation data in the Switchboard corpus. 
Inspired by the strategy proposed in Pepino et. al.~\cite{pepino2021emotion}, we fine-tune the transformer layers in the wav2vec2.0 network while keeping the lower convolutional layers unchanged. The hidden layer outputs from the wav2vec model, for all the transformer layers, are summed at the frame-level, and passed through a 1-D CNN network. 
  Finally, the embeddings from the CNN network are average pooled over the utterance level to generate audio embeddings for the given utterance.
 The embeddings obtained at the output of the 1D-CNN network are considered for the contextual GRU layer.
\subsubsection{Text embedding extractor}\label{textfeat}
The embedding extractor on the text data follows a similar architecture to that of the audio feature extraction. We obtain the word embeddings through a pre-trained RoBERTa model~\cite{liu2019roberta} and splice the last four hidden layer representations from the model. A bi-GRU layer allows the incorporation of intra-utterance context to generate utterance level embeddings.  Unlike the audio embedding  extraction, all the layers of the RoBERTa and bi-GRU model are fine-tuned. 
\subsection{HCAM Stage II}\label{stage2}
The utterance-level embeddings obtained from the previous modeling stage are used in this stage, where we introduce inter-utterance context by means of a bidirectional gated recurrent unit (Bi-GRU) architecture with self-attention mechanism. The representations  extracted from models trained on each individual utterance in stage I are further enhanced with the conversational context information. 
\subsubsection{Inter-utterance contextual GRU}\label{context}

 % ERC has a number of challenges like dependence on other speakers in the conversation, personality of the speaker or the listener and the current mood of the persons engaging in the conversation. In practice, however, we may not have information of the speaker identity of each utterance. Keeping this constraint in mind, 
We propose a simple block, called the contextual-GRU, which takes into account the information from all utterances in the conversation.  
The block diagram of the contextual GRU is shown in Fig. \ref{fig:context gru}.  While the bi-GRU itself can incorporate contextual information, it may not capture long term dependencies in the conversations that have a large number of utterances. For this reason, self-attention is used, which allows effective modeling of the long-term context. 
The output from the self-attention block is processed by a position wise feed forward layer with ReLU activation. 
%Once the models are trained for the three modalities of audio, text and ASR, we use the contextual embeddings of each utterance for the next stage of the model.
\subsection{HCAM Stage III}\label{stage3}
The third stage of the model consists of the effective fusion of the embeddings from the different modalities. 
%The context modeling layers, as described in Sec.\ref{stage2}, could have been trained in an end-end fashion with the fusion module. However, as will be shown later, the two modalities often show a large deviation in performance. In such cases, the better performing modality often suffers due to alignment with the weaker modality. This in turn leads to a drop in overall performance of the emotion recognition system. 
%The hierarchical training paradigm allows us to model the conversation context for each modality independently, with the audio and text embeddings   being combined by means of a co-attention network.

\subsubsection{Multi-modal fusion}\label{coatt}
\begin{figure}
    \centering
    \includegraphics[width=\columnwidth,trim={4cm 6cm 0cm 2cm},clip]{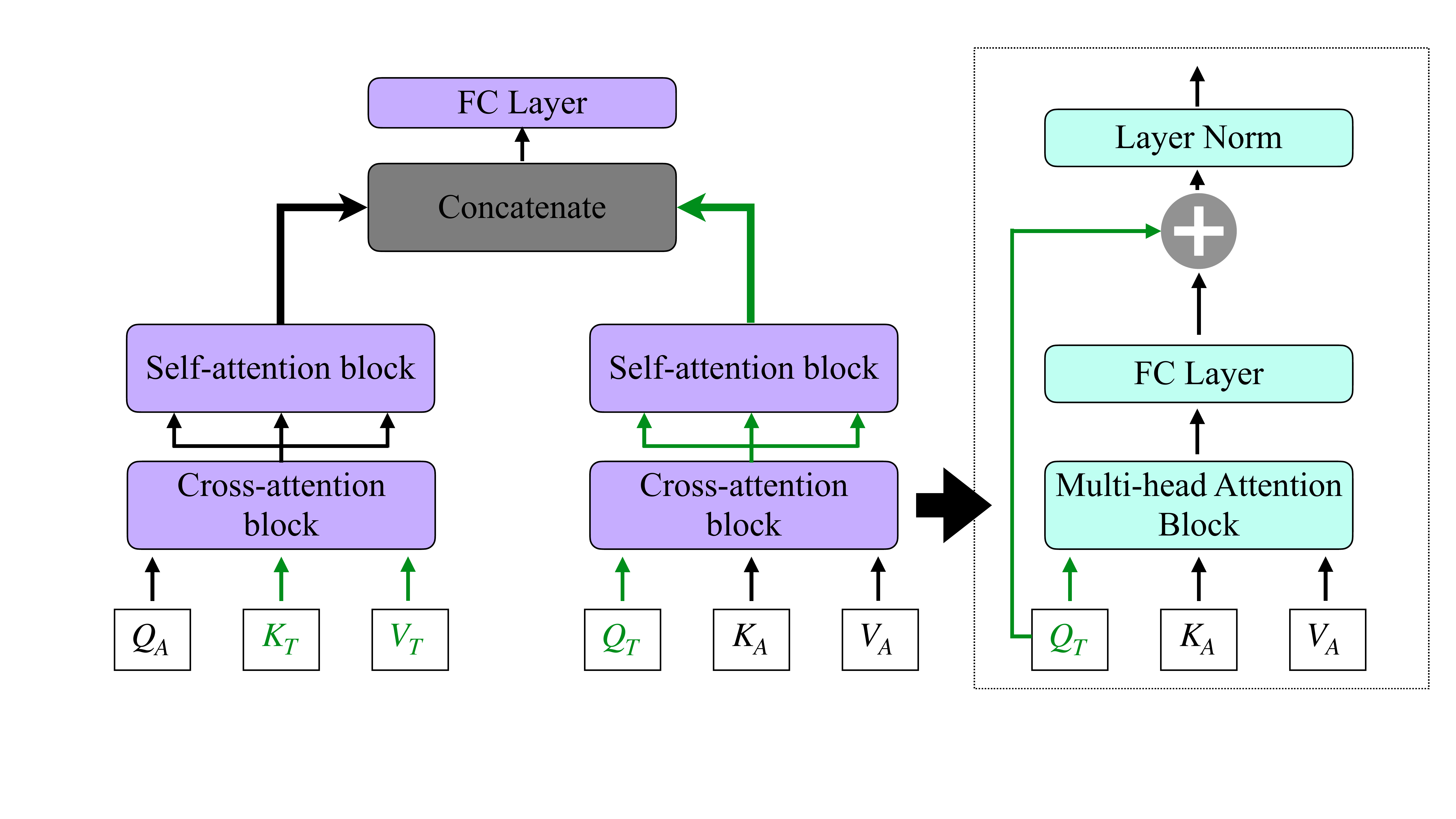}
    \caption{The co-attention network used in the proposed model. It consists of two sub-blocks - the cross-attention and the self-attention blocks.}
    \label{fig:attention}
\end{figure}

The network architecture of the co-attention block, is shown in Fig. \ref{fig:attention}, and it consists of two sub-blocks, namely cross-attention and self-attention. 
% \subsubsection{Cross Attention}
% The cross-attention block used in the model is shown in Fig. \ref{fig:attention}. The aim of this block is to enhance the representation of one modality (represented by $Q_A$ for example) with the help of the key and value from another modality (represented by $K_T$ and $V_T$ respectively for example). For this, we compute the cosine similarity between the representations of the two modalities,  $Q_A$ and $K_{T}$), and then take a weighted average of $V_{T}$ accordingly. The computation is shown in equations \ref{crossatt1}, \ref{crossatt2} and \ref{crossatt3}. 
% \begin{eqnarray}
% &Atten(Q_A, K_{T}, V_{T}) = softmax(\frac{Q_{A}K_{T}^{T}}{\sqrt{d_k}})V_{T}\label{crossatt1}\\
% &Q_A = Q_A + Atten(Q_A, K_{T}, V_{T})\label{crossatt2}\\
% &Q_A = LayerNorm(Q_A)\label{crossatt3}
% \end{eqnarray}
% We note that this computation helps us align two modalities and get a representation involving both of them.
The self-attention and cross attention schemes were discussed in Sec.~\ref{sec:self_cross_attn}. In the co-attention network, we exploit the cross-attention between the modalities. 
There are two ways of performing the cross-attention between audio and text, one where the query representations are derived from the audio data, while the key and value representations are derived from the text data.
The other way is the reversal of the roles of the audio and text. The cross-modal embeddings at the output of the cross-attention blocks are further enriched by adding a self-attention block.  The two arms of cross-attention, as shown in Fig.~\ref{fig:attention}, are concatenated and this forms the multi-modal representation. 

The final representations, which combines the information from audio and text, is passed through a position-wise feed forward layer. 
% We note that in certain cases, one of the two modalities may be considerably poorer in performance than the other. One such example would be when poor audio quality in the dataset degrades the performance of the audio based classifier as compared to the models trained on the provided text transcripts. The poor audio quality would lead to a poor performance of the pre-trained ASR system as well. Therefore, the performance of the more accurate modality (text in this case) may suffer due to the cross-modal interactions leading to poor performance of the overall recognition system. The representations obtained at the end of stage 3 of training for audio-ASR are further passed through a self-attention network. A separate self-attention network processes the embeddings obtained from the contextual GRU for the text transcripts. The output from these self-attention networks are further concatenated with that of the co-attention network between ASR-audio and text. These self-attention blocks though working on a single modality (audio-ASR or text) are trained jointly with the co-attention block in the final stage of our training schedule. 
% Position-wise feed forward networks are then used for predicting the emotion labels for each utterance in the conversation.\par

% The embeddings from this co-attention block (referred to as audio-ASR embeddings) are then combined with the embeddings from the uni-modal model trained on the provided text transcripts. This constitutes the final stage of our training procedure.

\subsection{Loss function}\label{loss}
\subsubsection{Supervised contrastive loss across conversations}\label{supconlossdescr}
%It is to be noted that the representations of the samples having same emotions across conversations should be more similar than samples having different emotions. The focal loss is aimed at classifying each utterance of the conversation correctly. This loss is not concerned with the utterance representations from different conversations.
We explore the supervised contrastive loss function, proposed by Khosla et.al~\cite{khosla2020supervised}, that encourages similarity between utterance representations from the same emotion class. Let us consider a mini-batch size of $\mathcal{B}$, where the utterance-level representations, derived from multiple conversations,  are denoted as $\{x_1, x_2, \dots, x_\mathcal{B}\}$. The features that appear in this loss formulation are normalized, that is, $||x_i|| = 1 \quad \forall i = \{1,2,\dots,\mathcal{B}\}$. We denote the corresponding labels as $\{y_1, y_2, \dots, y_\mathcal{B}\}$. Considering the sample with index $j$, the set of positive examples from the mini-batch is denoted by $P_j: \{i \in \mathcal{B} \quad s.t. \quad y_i = y_j\}$. The supervised contrastive loss is,
\begin{equation}\label{supconeq}
    L^{sup-con} = \sum_{j\in \mathcal{B}} \frac{-1}{|P_j|}\sum_{p \in P_j} log \frac{exp(x_j^T x_p/\tau)}{\sum \limits_{a\in \mathcal{B}}exp(x_j^T x_a/\tau)},
\end{equation}
where, $\tau$ is a hyper-parameter indicating the temperature of this loss.
Generally, contrastive losses have been used in representation learning tasks. The dependence of the contrastive loss on the ground truth labels enables one to use these losses in supervised settings too. In addition to the cross-entropy loss, this loss makes the system focus on the hard-to-classify samples. However, unlike in representation learning tasks, the datasets for ERC are not large enough to be able to classify the utterances by means of a contrastive loss alone. We therefore, use the supervised contrastive loss in combination with the cross-entropy loss for training our emotion recognition modules.
\subsubsection{Combined loss}
%In problems employing the supervised contrastive loss, authors generally use a pre-training step to train the models. The small size of ERC datasets however does not allow to follow such a training paradigm. 
We use a convex combination of the the cross entropy loss and the supervised contrastive loss. The final loss, used to train all the stages of our model, is given by,
\begin{equation}\label{lossfinal}
    L^{tot} = \beta CE(y, \hat{y}) + (1-\beta)L^{sup-con},
\end{equation}
where $\beta$ is a hyper-parameter in the range of $[0,1]$.
\subsection{Inference}\label{inference}
During the inference in a particular stage, we combine the predictions of the model from the previous stage. Thus, once the contextual GRU model is trained in stage II, we combine the predictions of the audio and text models from stage I. During the inference in stage III, after fusion of the audio and text representations, we combine the predictions of the contextual GRU model (for audio and text separately). Let the softmax outputs for an utterance $x$ at the end of stage I be $p_a^{1}$ (for audio) and $p_t^{1}$ (for text). These are combined with the unimodal contextual GRU predictions (denoted by $p_a^{2}$ and $p_t^2$), with weights denoted by $\alpha_{a}^{1\rightarrow2}$ and $\alpha_{t}^{1\rightarrow2}$, as follows,
\begin{eqnarray}
&\hat{y}_{a}^2(x) = \alpha_a^{1\rightarrow2} p_a^{2}(x) + (1-\alpha_a^{1\rightarrow2}) p_a^{1}(x)\label{aud12}\\
&\hat{y}_{t}^2(x) = \alpha_t^{1\rightarrow2} p_t^{2}(x) + (1-\alpha_t^{1\rightarrow2}) p_t^{1}(x)\label{text12}
\end{eqnarray}
Similarly, we combine the outputs from stage II with the predictions from the co-attention module ($p_c$) as follows,
\begin{equation}\label{predfinal}
    \hat{y}(x) = \alpha_c p_c(x) + \alpha_a^{2\rightarrow3} \hat{y}_{a}^2(x) + \alpha_t^{2\rightarrow3} \hat{y}_{t}^2(x)
\end{equation}

where $ \hat{y}(x)$ refers to the  final predictions used for  classification. All the combination weights used in equations (\ref{aud12}), (\ref{text12}) and (\ref{predfinal}) are decided based on validation set performance.
\section{Experiments and Results}\label{experiments}
% In this section, we introduce the different conversational datasets used followed by the experimental setup for them. 
\subsection{Datasets}
We evaluate our work on three widely used datasets, IEMOCAP~\cite{busso2008iemocap}, MELD~\cite{poria2019meld} and CMU-MOSI~\cite{zadeh2016mosi}. Unlike the first two, CMU-MOSI is not multi-speaker  in nature as it has single speaker monologues. Further, we use only text and audio modalities for emotion recognition task in this paper.
\subsubsection{IEMOCAP}
The IEMOCAP dataset consists of $151$ video recordings split into 5 sessions. Each of these sessions is a conversation between a pair of people, one male and one female. Each recording is split into multiple utterances. There are a total of $10,039$  utterances, each of which is labeled by human annotators as belonging to one of the $10$ emotions -  angry, happy, sad, neutral, frustrated, excited, fearful, surprised, disgusted or ``other''. Keeping in line with previous works, we do a four-way classification task where we consider angry, happy, sad, neutral and excited categories (with excited and happy categories merged). We have a total of $5531$ utterances from the four emotion labels. We also have a separate setting of $6$ emotional classes,  as has been done in some of the prior works such as~\cite{majumder2019dialoguernn}. The first $6$ emotion classes are considered resulting in a total of $7433$ utterances. The dataset is imbalanced with the least number of samples for the happy emotion ($648$ utterances during training). For both these cases, we consider session $5$ for testing purposes. We choose session $1$ for validating our models and sessions $2-4$ for training.

\subsubsection{MELD}
The MELD dataset is a multi-party dataset created from video clippings of the popular TV show, ``Friends''. The training data consists of $9988$ utterances, validation data consists of $1108$ utterances and  test data consists of $2610$ utterances.
%Unlike IEMOCAP, the test setting is not speaker independent. 
A seven way classification task is performed on this dataset, with each utterance being labeled as one of the $7$ emotions - angry, sad, joy, neutral, fear, surprise or disgust. Like the IEMOCAP 6-way classification problem, this is also an imbalanced dataset with neutral being the most dominant class label and disgust being the least frequent label ($4710$ and $271$ training utterances respectively).
\subsubsection{CMU-MOSI}
The CMU-MOSI dataset has a total of $93$ monologues divided into $2199$ utterances. Each monologue is divided into several utterances and is labeled in the range of $[-3, 3]$. Following previous works, we treat this as a binary classification problem with utterances having sentiment values in the range $[-3, 0)$ being classified as negative sentiment and those with values in the range $[0, 3]$ as positive sentiment. For dataset partitioning, we follow the prior work by Poria et. al~\cite{poria2017context}, where the first  $62$ monologues are used for training and validation while the last $31$ monologues are used for testing. Of the $62$ monologues, we use $49$ for training our model and the rest $13$ for validation.
\begin{table}[t!]
\centering
\caption{Results on the different datasets in terms of weighted F1-score.}\label{tab:results}
    \vspace{-0.0in}

\resizebox{\columnwidth}{!}{%
\begin{tabular}{@{}l|cc|c|c@{}}
\toprule
\multirow{2}{*}{Modality (Training Stage)} & \multicolumn{2}{c|}{IEMOCAP} & \multirow{2}{*}{MELD} & \multirow{2}{*}{CMU-MOSI} \\ \cmidrule(lr){2-3}
 & \multicolumn{1}{c|}{$4$ way} & $6$ way &  &  \\ \midrule
Audio (Stage I) & \multicolumn{1}{c|}{$64.3\%$} & $55.6\%$ & $48.2\%$ & $64.4\%$ \\
Audio (Stage II) & \multicolumn{1}{c|}{$78.7\%$} & $65.7\%$ & $50.1\%$ & $67.4\%$ \\ \midrule
Text (Stage I) & \multicolumn{1}{c|}{$68.4\%$} & $53.8\%$ & $63.3\%$ & $84.3\%$ \\
Text (Stage II) & \multicolumn{1}{c|}{$81.4\%$} & $64.4\%$ & $65.6\%$ & $85.4\%$ \\ \midrule
Audio + Text (Stage III) & \multicolumn{1}{c|}{$85.9\%$} & $70.5\%$ & $65.8\%$ & $85.8\%$ \\ \bottomrule
\end{tabular}
    \vspace{-0.1in}

}

\end{table}
\subsection{Implementation details}\label{implement}
%The proposed architecture is trained in a hierarchical fashion. The text and audio embedding extractor networks are trained using emotion class labels at the utterance-level (Stage 1). Following this training, the contextual GRU model is trained using all the utterances in a given conversation (Stage 2). Further, the co-attention layer of audio with ASR transcripts is trained (Stage 3) and subsequently, the rest of the network layers in the model (co-attention between audio-ASR and text, self-attention and fully connected layers) are trained using the emotion labels for all the utterances in the entire conversation (Stage 4). 

%The MELD and the 6-way classification of IEMOCAP being imbalanced datasets, are trained with the parameter $\gamma = 2$ in the expression for FoCal loss (Eq.~\ref{focal}). The other two datasets of CMU-MOSI and IEMOCAP 4-way classification, being more balanced classification problems, are trained with the cross-entropy loss. The value of this hyper-parameter $\gamma$ was decided based on the performance on the validation data. We further investigate the choice of $\beta$, as defined in Eq.(\ref{lossfinal}) and the value of the temperature parameter in the supervised contrastive loss (Eq.\ref{supconeq}).
The models are trained with Adam optimizer using a learning rate of $1e-5$ and a batch size of $32$ in MELD and IEMOCAP dataset, while the batch size is reduced to $8$ for the CMU-MOSI dataset. All the experiments reported in this work use $5$ random weight initialization choices. The mean performance using the random initializations are reported in all the experiments below. We run the different stages of our model for $100$ epochs as we found the validation performance to saturate within this limit. We also employ gradient clipping with a L2 norm of $0.25$ in all our implementation.

\subsection{Results}
%Our proposed model, as discussed before, uses two modalities, namely audio and text. As outlined in Section \ref{proposal}, we enhance the audio modality by incorporating information from ASR transcripts. In this section we mention the results that we have obtained for the three datasets that we have used. Since our model is trained in a hierarchical manner, 
We report the performance of the proposed model for each individual modality followed by the performance of the model on the multi-modal setting.  The key results on the three datasets are shown in Table \ref{tab:results}. 

The following are the observations from these results, 
\begin{enumerate}[label=(\roman*)] 
    \item In the IEMOCAP dataset, the audio and the text modalities perform relatively similarly, while in the MELD and CMU-MOSI datasets, the audio results in an inferior performance compared to the text domain
    %\item The speech based emotion recognition (combining audio with ASR transcripts) results in substantial improvements over either of the two inputs individually. We observe relative improvements of $21.5$\%, $15.1$\%, $4.6$\% and $1.9$\% for the co-attention based combination of audio+ASR over either speech or ASR output for the IEMOCAP 4-way, IEMOCAP 6-way, MELD and CMU-MOSI datasets, respectively. 
    \item The context addition framework proves to be effective for all the three datasets. The proposed contextual GRU architecture with self-attention, though a simple architecture, leads to an improvement for both the modalities. For audio, the relative improvement over the stage I performance is $40.3\%$, $22.7\%$, $3.7\%$ and $8.4\%$ for IEMOCAP 4-way, IEMOCAP 6-way, MELD and CMU-MOSI datasets, respectively. Similarly, for the textual modality, we notice a relative improvement of $41.1\%$, $22.9\%$, $6.3\%$ and $7\%$ respectively. IEMOCAP has the largest number of utterances per conversation among the three datasets while MELD has the lowest. As the number of utterances increase in the datasets, the contextual modeling becomes more effective
    \item The final multi-modal fusion achieves the best performance on all the three datasets over any of the individual modality. The relative improvements of $24.2$\%, $14$\%, $0.6$\% and $2.7$\% are observed for multi-modal results over the best individual modality in the IEMOCAP 4-way, IEMOCAP 6-way, MELD and CMU-MOSI datasets, respectively. These results show that, even when some modalities are inferior to the others in the emotion classification task, the co-attention mechanism is able to effectively improve over the best individual modality
\end{enumerate}
\section{Discussion}\label{discuss}
\begin{table}[t!]
\caption{\label{otherresultsiemocap}Comparison with other works for IEMOCAP 4-way classification. All scores are the weighted F1 scores.}
    \vspace{-0.1in}
\begin{center}
\resizebox{0.8\columnwidth}{!}{%
\begin{tabular}{@{}l|cl|c|c@{}}
\toprule
\multicolumn{1}{c|}{System} & \multicolumn{2}{c|}{Audio} & Text & \begin{tabular}[c]{@{}c@{}}Audio+Text\end{tabular} \\ \midrule
Majumder et al.~\cite{majumder2018multimodal} & \multicolumn{2}{c|}{$57.1\%$} & $73.6\%$ & $76.1\%$ \\
Mai et al.~\cite{mai2019divide} & \multicolumn{2}{c|}{$38.2\%$} & $\mathbf{81.5\%}$ & $80.6\%$ \\
Li et al.~\cite{li2019towards} & \multicolumn{2}{c|}{$69.3\%$}  & - & $79.1\%$ \\
Mittal et al.~\cite{mittal2020m3er} & \multicolumn{2}{c|}{-}  & - & $82.4\%$ \\
Mai et al.~\cite{mai2019locally} & \multicolumn{2}{c|}{-}  & - & $82.5\%$ \\
Dutta et al.~\cite{dutta2022multimodal} & \multicolumn{2}{c|}{$73.5\%$}  & $78.9\%$ & $83.7\%$ \\
Lian et al.~\cite{lian2022smin}& \multicolumn{2}{c|}{-}  & - & $84.8\%$ \\\hline
HCAM (this work) & \multicolumn{2}{c|}{$\mathbf{78.7\%}$}  & $81.4\%$ & $\mathbf{85.9\%}$ \\ \bottomrule
\end{tabular}%
}
%    \vspace{-0.1in}

\end{center}
\end{table}

\begin{table}[t!]
\caption{\label{otherresultsiemocap6}Comparison with other works for IEMOCAP 6-way classification. All scores are the weighted F1 scores. * indicates our implementation.}
    \vspace{-0.1in}

\begin{center}
\resizebox{0.8\columnwidth}{!}{%
\begin{tabular}{@{}l|cl|c|c@{}}
\toprule
\multicolumn{1}{c|}{System} & \multicolumn{2}{c|}{Audio} & Text & \begin{tabular}[c]{@{}c@{}}Audio+Text\end{tabular} \\ \midrule
Majumder et al.~\cite{majumder2019dialoguernn} & \multicolumn{2}{c|}{-} &  - & $62.8\%$ \\
Ghosal et al.~\cite{ghosal2019dialoguegcn} & \multicolumn{2}{c|}{-}  & - & $64.2\%$ \\
Shen et al.~\cite{shen2021dialogxl} & \multicolumn{2}{c|}{-}  & - & $65.9\%$ \\
Mao et al.~\cite{mao2021dialoguetrm} & \multicolumn{2}{c|}{-} & - & $69.7\%$ \\
Lian et al.~\cite{lian2022smin}& \multicolumn{2}{c|}{-}  & - & $70.5\%$ \\
Li et al.~\cite{li-etal-2022-emocaps}& \multicolumn{2}{c|}{-}  & - & $69.3\%^{*}$ \\\hline
HCAM (this work) & \multicolumn{2}{c|}{$65.7\%$}  & $64.4\%$ & $\mathbf{70.5\%}$ \\ \bottomrule
\end{tabular}%
}
 %   \vspace{-0.2in}

\end{center}
\end{table}

\subsection{Comparison with other works}
We present a comparative study of our work along with other existing works in the literature.
%As our model enhances the outputs from the speech signals with the ASR transcripts, we consider the output from our Audio+ASR system as the performance on audio signals alone. \par 
The comparison with the relevant works are shown in Table \ref{otherresultsiemocap} for the IEMOCAP dataset with 4-way classification. We see that the performance of our model on audio signals is significantly better than other works (by a relative margin of about $20\%$). This is attributed to the efficient contextual modeling of the audio representations from the wav2vec model.

While we do not improve the state-of-the-art results in the text modality, we achieve state-of-the-art results on the multi-modal fusion task by a relative margin of $7\%$ over the previous best result. We also show a comparison with prior works reporting on the 6-way classification in the IEMOCAP dataset (Table \ref{otherresultsiemocap6}). For this unbalanced classification setting, we notice that our model matches the results reported by Lian et al~\cite{lian2022smin}. 

We compare with other relevant works on the MELD dataset in Table \ref{otherresultsmeld}. Similar to IEMOCAP, we notice a considerable improvement in the audio only performance. We improve the current state-of-the-art results in audio signal performance by a relative margin of approximately $14\%$. We improve the performance of our model on textual modality alone as well, where we achieve a relative improvement of $8.5\%$. Subsequently we get a relative improvement of approximately $4\%$ after the fusion module. It is noteworthy that, MELD is the most imbalanced of the three datasets with maximum number of speakers in a conversation. Our model, however, does not depend on any speaker information for emotion modeling unlike some prior works~\cite{mao2021dialoguetrm}. 
%An efficient modeling of the speaker information would lead to further improvements in the emotion recognition performance.
%The inability of the ASR embeddings to enhance the audio classifier results in a considerable difference in the performance of the audio and text modality, leading to only a small improvement upon fusion of text and audio.\par

In Table \ref{otherresultsmosi}, we compare our proposed model with other works on the CMU-MOSI dataset. Due to the small size of the dataset (only $2199$ utterances), the proposed model is prone to overfitting in this dataset. However, we note that the performance of the model in the audio modality is comparable to the previously reported best results on audio inputs alone. We improve upon the best reported text-only performance by a relative margin of $30\%$, while the model working on audio and text together improves upon the previous state-of-the-art performance by a relative margin of $26\%$.
\begin{table}[t!]
\caption{\label{otherresultsmeld}Comparison (in terms of weighted F1 scores) with prior works on the MELD dataset.}
    \vspace{-0.1in}

\begin{center}
\resizebox{0.8\columnwidth}{!}{%
\begin{tabular}{@{}l|cl|c|c@{}}
\toprule
\multicolumn{1}{c|}{System} & \multicolumn{2}{c|}{Audio} & Text & \begin{tabular}[c]{@{}c@{}}Audio+Text\end{tabular} \\ \midrule
Poria et al.~\cite{poria2017context} & \multicolumn{2}{c|}{$39.1\%$} & $56.4\%$ & $59.3\%$ \\
Majumder et al.~\cite{majumder2019dialoguernn} & \multicolumn{2}{c|}{$41.8\%$} & $57\%$ & $60.3\%$ \\
Zhang et al.~\cite{zhang2019modeling} & \multicolumn{2}{c|}{$42.2\%$}  & $57.4\%$ & $59.4\%$ \\
Ghosal et al.~\cite{ghosal2019dialoguegcn} & \multicolumn{2}{c|}{-}  & $58.1\%$ & - \\
Li et al.~\cite{li2020hitrans} & \multicolumn{2}{c|}{-} & $61.9\%$ & - \\
Shen et al.~\cite{shen2021dialogxl} & \multicolumn{2}{c|}{-} & $62.4\%$ & - \\
Mao et.al~\cite{mao2021dialoguetrm} & \multicolumn{2}{c|}{-}  & - &$63.6\%$ \\ 
Lian et al.~\cite{lian2022smin}& \multicolumn{2}{c|}{-}  & - & $64.5\%$ \\
Li et al.~\cite{li-etal-2022-emocaps}& \multicolumn{2}{c|}{-}  & - & $64\%$ \\\hline
HCAM (this work) & \multicolumn{2}{c|}{$\mathbf{50.1\%}$} & $\mathbf{65.6\%}$ & $\mathbf{65.8\%}$ \\ \bottomrule
\end{tabular}%
}
%\vspace{-0.2in} 
\end{center}
\end{table}

\begin{table}[t!]
\caption{\label{otherresultsmosi}Comparison with other works for the CMU-MOSI dataset. All scores are the weighted F1 scores.}
    \vspace{-0.1in}

\begin{center}
\resizebox{0.8\columnwidth}{!}{%
\begin{tabular}{@{}l|cl|c|c@{}}
\toprule
\multicolumn{1}{c|}{System} & \multicolumn{2}{c|}{Audio}  & Text & \begin{tabular}[c]{@{}c@{}}Audio+Text\end{tabular} \\ \midrule
Poria et al.~\cite{poria2017multi} & \multicolumn{2}{c|}{$60.1\%$}  & $79.1\%$ & $80.1\%$ \\
Zadeh et al.~\cite{zadeh2017tensor} & \multicolumn{2}{c|}{$67.4\%$} & $75.6\%$ & - \\
Chen et al.~\cite{chen2017multimodal} & \multicolumn{2}{c|}{-} & $67.3\%$ & - \\ 
Lian et al.~\cite{lian2022smin}& \multicolumn{2}{c|}{-}  & - & $80.8\%$ \\\hline
HCAM (this work) & \multicolumn{2}{c|}{$\mathbf{67.4\%}$}  & $\mathbf{85.4\%}$ & $\mathbf{85.8\%}$\\ \bottomrule
\end{tabular}%
}
\end{center}
\end{table}
\begin{table}[t!]
\caption{\label{hierarchy}Weighted F1 score of our system with different training paradigms}
\vspace{-0.1in}
\begin{center}
\resizebox{0.9\columnwidth}{!}{
    \begin{tabular}{@{}c|cc|c|c@{}}
\toprule
\multirow{2}{*}{\begin{tabular}[c]{@{}l@{}}Training Paradigm\end{tabular}} & \multicolumn{2}{c|}{IEMOCAP} & \multirow{2}{*}{MELD} & \multirow{2}{*}{CMU-MOSI} \\ \cmidrule(lr){2-3}
 & \multicolumn{1}{c|}{4 way} & 6 way &  &  \\ \midrule
Hierarchical & \multicolumn{1}{c|}{$85.9\%$} & $70.5\%$ & $65.8\%$ & $85.8\%$ \\
Non-hierarchical & \multicolumn{1}{c|}{$82.3\%$} & $68.5\%$ & $65\%$ & $84.4\%$ \\ \bottomrule
\end{tabular}}
\end{center}
\vspace{-0.1in}
\end{table}

\begin{table}[t!]
\caption{\label{tab:self-atten}Weighted F1 score of our system for the two modalities when the self-attention block is removed from the contextual GRU.}
\begin{center}
\resizebox{\columnwidth}{!}{
\begin{tabular}{@{}l|c|cc|c|c@{}}
\toprule
\multirow{2}{*}{Modality} & \multirow{2}{*}{Stage II model} & \multicolumn{2}{c|}{IEMOCAP} & \multirow{2}{*}{MELD} & \multirow{2}{*}{CMU-MOSI} \\ \cmidrule(lr){3-4}
 &  & \multicolumn{1}{l|}{4-way} & \multicolumn{1}{l|}{6-way} &  &  \\ \midrule
\multirow{2}{*}{Audio} & \begin{tabular}[c]{@{}c@{}}Bi-GRU without\\ self-attention\end{tabular} & \multicolumn{1}{c|}{$75.2\%$} & $63.8\%$ & $50.1\%$ & $67\%$ \\\cmidrule(lr){2-2}
 & \begin{tabular}[c]{@{}c@{}}Bi-GRU with\\ self-attention\end{tabular} & \multicolumn{1}{c|}{$78.7\%$} & $65.7\%$ & $50.1\%$ & $67.4\%$ \\ \midrule
\multirow{2}{*}{Text} & \begin{tabular}[c]{@{}c@{}}Bi-GRU without\\ self-attention\end{tabular} & \multicolumn{1}{c|}{$79.9\%$} & $62.5\%$ & $65\%$ & $85.8\%$ \\\cmidrule(lr){2-2}
 & \multicolumn{1}{c|}{\begin{tabular}[c]{@{}c@{}}Bi-GRU with \\ self-attention\end{tabular}} & \multicolumn{1}{c|}{$81.4\%$} & $64.4\%$ & $65.6\%$ & $85.4\%$ \\ \bottomrule
\end{tabular}}
\end{center}
\end{table}
\subsection{Impact of hierarchical modeling}
In order to understand the advantages of our hierarchical modeling approach, we modify our training paradigm where we combine stages II  and III of training.  The results for these modifications in the training paradigm are shown in Table \ref{hierarchy}. We note that, for all the four dataset settings, we achieve a better performance with the proposed hierarchical modeling with no change in the model architecture. These experiments highlight the benefits of a curriculum style design of the stages proposed in the HCAM framework.

\subsection{Importance of self-attention in stage II}
In order to understand the role of self-attention  in the contextual GRU, we run an ablation experiment where we remove the self-attention block from the contextual GRU. The results from this experiment are shown in Table~\ref{tab:self-atten}. We note that self-attention improves the performance for both the modalities  in IEMOCAP. This is expected as the conversation length is more in the case of IEMOCAP (conversation length of $110$ utterances). In a departure from the other datasets, we see an absolute drop of $0.4\%$ for the textual modality for CMU-MOSI dataset on the introduction of the self-attention block. This may partly be due to the small size of this dataset, which leads the Bi-GRU model to overfit.
\begin{table*}[ht!]
\centering
\caption{The Word Error Rate (WER) (\%) for the Google ASR system on the three datasets used. Abbreviations used: Happy:\textit{Hap.}, Neutral:\textit{Neu.}, Angry:\textit{Ang.}, Excited:\textit{Exc.}, Frustrated:\textit{Fru.}, Disgust:\textit{Dis.}, Positive:\textit{Pos.} and Negative:\textit{Neg.}}\label{tab:asr_wer}%
\resizebox{\textwidth}{!}{%
\begin{tabular}{@{}l|ccclll|lllllll|ll@{}}
\toprule
\multirow{2}{*}{\begin{tabular}[c]{@{}l@{}}Dataset\\ Splits\end{tabular}} & \multicolumn{6}{c|}{IEMOCAP} & \multicolumn{7}{c|}{MELD} & \multicolumn{2}{c}{CMU-MOSI} \\ \cmidrule(l){2-16} 
 & \multicolumn{1}{l|}{Hap.} & \multicolumn{1}{l|}{Sad} & \multicolumn{1}{l|}{Neu.} & \multicolumn{1}{l|}{Ang.} & \multicolumn{1}{l|}{Exc.} & Fru. & \multicolumn{1}{l|}{Ang.} & \multicolumn{1}{l|}{Sad} & \multicolumn{1}{l|}{Neu.} & \multicolumn{1}{l|}{Fear} & \multicolumn{1}{l|}{Sur.} & \multicolumn{1}{l|}{Dis.} & Joy & \multicolumn{1}{l|}{Pos.} & Neg. \\ \midrule
Train & \multicolumn{1}{c|}{\multirow{2}{*}{36{}{}}} & \multicolumn{1}{c|}{\multirow{2}{*}{44{}{}}} & \multicolumn{1}{c|}{\multirow{2}{*}{31.2{}{}}} & \multicolumn{1}{l|}{\multirow{2}{*}{23.4{}{}}} & \multicolumn{1}{l|}{\multirow{2}{*}{36.5{}{}}} & \multirow{2}{*}{28{}{}} & \multicolumn{1}{l|}{50.1} & \multicolumn{1}{l|}{50.1} & \multicolumn{1}{l|}{50.2} & \multicolumn{1}{l|}{54.8} & \multicolumn{1}{l|}{59} & \multicolumn{1}{l|}{47.5} & 54.1 & \multicolumn{1}{l|}{37.3} & 37.9 \\
Val & \multicolumn{1}{c|}{} & \multicolumn{1}{c|}{} & \multicolumn{1}{c|}{} & \multicolumn{1}{l|}{} & \multicolumn{1}{l|}{} &  & \multicolumn{1}{l|}{47.4} & \multicolumn{1}{l|}{49.9} & \multicolumn{1}{l|}{46.4} & \multicolumn{1}{l|}{52.8} & \multicolumn{1}{l|}{58.8} & \multicolumn{1}{l|}{46.6} & 50.7 & \multicolumn{1}{l|}{34} & 37.9 \\
Test & \multicolumn{1}{c|}{34.5{}{}} & \multicolumn{1}{c|}{47.5{}{}} & \multicolumn{1}{c|}{30.2{}{}} & \multicolumn{1}{l|}{27.3{}{}} & \multicolumn{1}{l|}{36{}{}} & 30.5{}{} & \multicolumn{1}{l|}{50.5} & \multicolumn{1}{l|}{49.2} & \multicolumn{1}{l|}{48.1} & \multicolumn{1}{l|}{59.5} & \multicolumn{1}{l|}{61.2} & \multicolumn{1}{l|}{49.3} & 54.5 & \multicolumn{1}{l|}{38.2} & 40.4 \\ \bottomrule
\end{tabular}
}
\end{table*}
\begin{figure*}[t!]
    \centering
    \includegraphics[width=\textwidth,trim={0cm 4.5cm 0cm 4cm},clip]{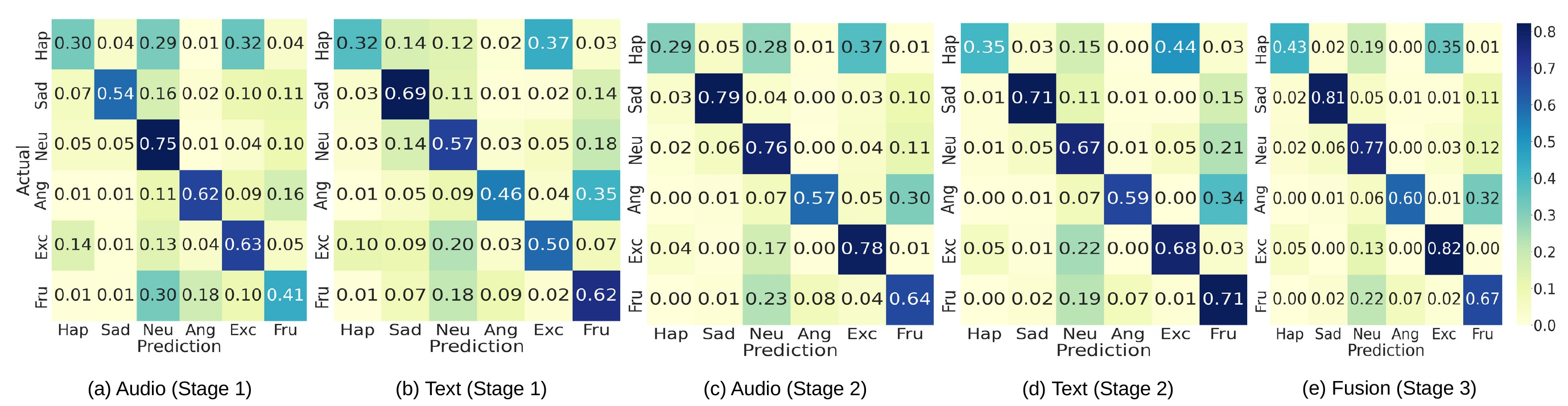}
        \vspace{-0.1in}
    \caption{Confusion matrices for the different stages of our model when run on IEMOCAP dataset with 6 classes. Abbreviations used: Happy:\textit{Hap.}, Neutral:\textit{Neu.}, Angry:\textit{Ang.}, Excited:\textit{Exc.}, Frustrated:\textit{Fru.}}
    \label{fig:conf}
  %  \vspace{-0.2in}
\end{figure*}
\subsection{Testing with ASR generated transcripts}
%The use of ASR transcripts as another modality for this particular problem gives considerable gains as shown in Table~\ref{tab:results}.
%As the ASR transcripts are used in an audio only setting, we do not train any ASR model to detect emotional speech. 
In order to understand the robustness of our model to noise in the text modality, we test our model with ASR transcripts in place of the provided transcripts. The training of the model is not modified from the previous experiments
%, however, during the testing phase, the provided text transcripts are replaced by the ASR generated transcriptions.

We use a pre-trained ASR system \footnote{\url{https://cloud.google.com/speech-to-text}\label{googlefoot}} for providing the transcripts. The word error rate (WER) of this ASR system  is reported  in Table \ref{tab:asr_wer} for each of the three datasets. As seen here, the WER on emotional conversational speech is significantly higher than those seen on other controlled datasets. The ASR performance in the case of MELD is the lowest, partly due to the high levels of background noise in the dataset.
\begin{table}[t!]
\caption{\label{tab:new_asr}Weighted F1 score of our system with ASR transcripts during test time on the datasets.}
\begin{center}
\resizebox{0.9\columnwidth}{!}{
\begin{tabular}{@{}l|cc|c|c@{}}
\toprule
\multicolumn{1}{c|}{\multirow{2}{*}{\begin{tabular}[c]{@{}c@{}}Text Transcripts\\ using test time\end{tabular}}} & \multicolumn{2}{c|}{IEMOCAP} & \multirow{2}{*}{MELD} & \multirow{2}{*}{CMU-MOSI} \\ \cmidrule(lr){2-3}
\multicolumn{1}{c|}{} & \multicolumn{1}{l|}{4-way} & \multicolumn{1}{l|}{6-way} &  &  \\ \midrule
ASR & \multicolumn{1}{c|}{$84.6\%$} & $68.1\%$ & $50.2\%$ & $80.1\%$ \\
Original & \multicolumn{1}{c|}{$85.9\%$} & $70.5\%$ & $65.8\%$ & $85.8\%$\\ \hline
Previous SOTA~\cite{dutta2022multimodal} & \multicolumn{1}{c|}{$77.3\%$} & - & - & - \\ \bottomrule
\end{tabular}}
\end{center}
\end{table}
In spite of the high WER, the model performs well on the IEMOCAP and CMU-MOSI datasets (with an absolute drop of $2.4\%$ for IEMOCAP $6$-way classification and $5.7\%$ for the CMU-MOSI) as shown in Table \ref{tab:new_asr}. For the baseline result in this setting, we use our previous work~\cite{dutta2022multimodal}, as we have not found a similar inference setting elsewhere.
%Hence out of the 4 test settings, we compare with our prior work for IEMOCAP $4$-way classification while we establish baselines for the other $3$. 
We achieve a significant relative improvement of $32\%$ over our previous work.

\begin{table}[t!]
\caption{The weighted F1 scores (in $\%$) with and without the supervised contrastive loss. }
\label{supcon}
\begin{center}
\resizebox{0.85\columnwidth}{!}{%
\begin{tabular}{@{}l|cc|c|c@{}}
\toprule
\multicolumn{1}{l|}{\multirow{2}{*}{Loss}} & \multicolumn{2}{c|}{IEMOCAP} & \multicolumn{1}{l|}{\multirow{2}{*}{MELD}} & \multicolumn{1}{l}{\multirow{2}{*}{CMU-MOSI}} \\ \cmidrule(lr){2-3}
\multicolumn{1}{c|}{} & 4-way & 6-way & \multicolumn{1}{l|}{} & \multicolumn{1}{l}{} \\ \midrule
CE & $84.5\%$ & $69.9\%$ & $65.5\%$  & $84.7\%$ \\
CE+Supcon & $85.9\%$ & $70.5\%$  & $65.8\%$  & $85.8\%$ \\
\bottomrule
\end{tabular}%
}
\end{center}
\end{table}
% \begin{table}[t!]
% \caption{The weighted F1 scores for the two imbalanced datasets are reported for the three uni-modal streams and for different values of $\gamma$ (Eq. (\ref{focal}))}
% \label{tab:gamma}
% \centering
% \resizebox{0.9\columnwidth}{!}{%
% \begin{tabular}{@{}c|ccc|ccc@{}}
% \toprule
% \multirow{2}{*}{$\gamma$} & \multicolumn{3}{c|}{IEMOCAP 6-way} & \multicolumn{3}{c}{MELD} \\ \cmidrule(l){2-7} 
%  & Audio & Text & ASR & Audio & Text & ASR \\ \midrule
% 0 & 60.3\% & 62.3\% & 60.1\% & 45.2\% & 61.5\% & 46\% \\
% 2 & 62.8\% & 62.8\% & 60.4\% & 48.1\% & 61.7\% & 46.2\% \\
% 3 & 62.7\% & 62.9\% & 60.3\% & 47.8\% & 61.7\% & 46.2\% \\ \bottomrule
% \end{tabular}%
% }
% \end{table}
\begin{figure}[t!]
    \centering
    \includegraphics[width=\columnwidth,trim={7cm 0cm 15cm 1cm},clip]{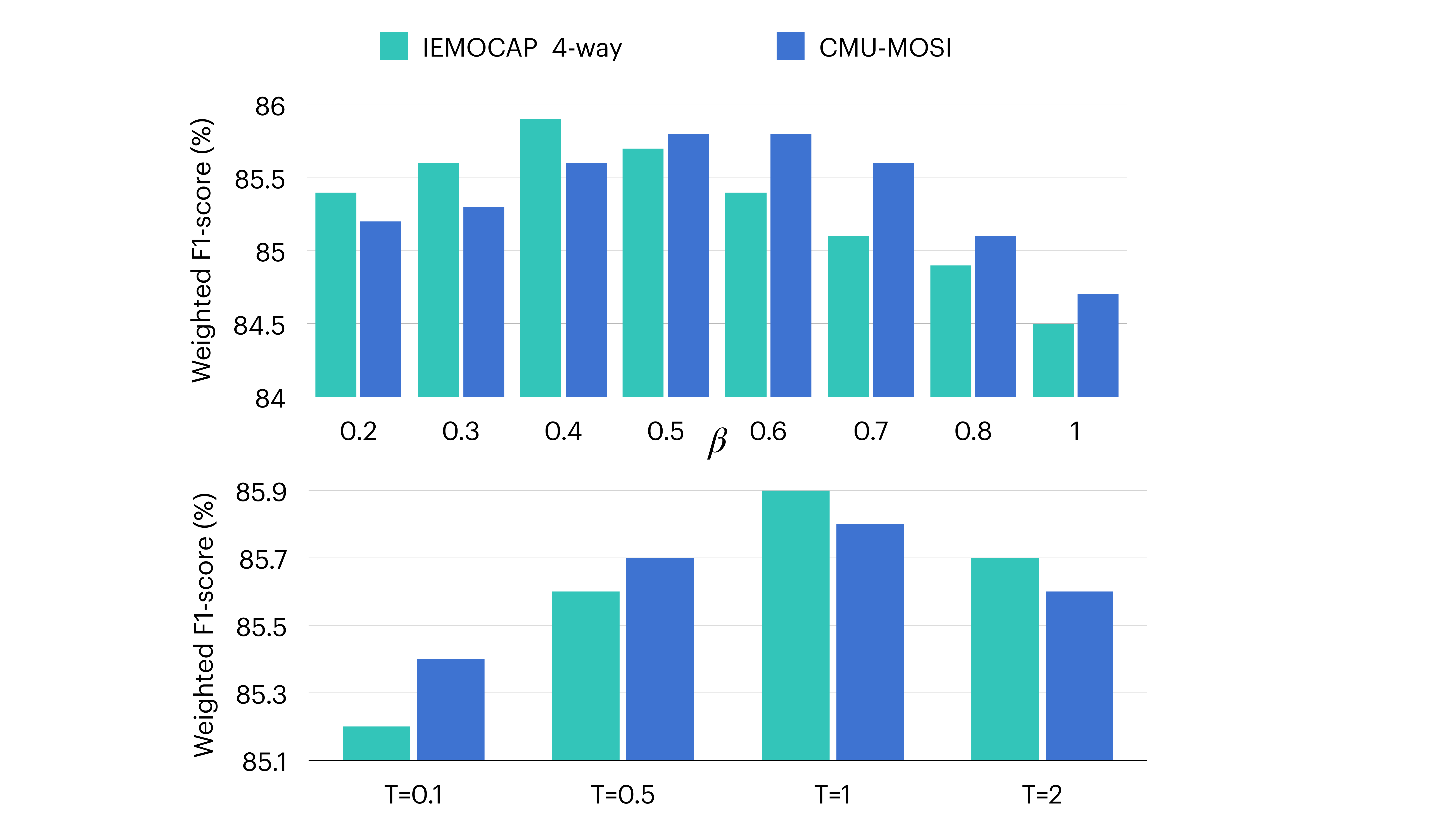}
    \caption{Variation of the test performance with change in $\beta$ (Eq.\ref{lossfinal}) and temperature parameter in the sup-con loss for IEMOCAP $4$-way and CMU-MOSI datasets.}
    \label{fig:supcon}
\end{figure}
\subsection{Role of the supervised contrastive loss}
The performance of the proposed model is evaluated when the training is performed without and with the supervised contrastive loss. These results are shown in Table \ref{supcon}. For all the datasets, we observe an improvement in the performance of our model with the inclusion of the supervised loss.

However, the performance improvements from   the supervised contrastive loss  varies from one dataset to the other.  
%The model learning therefore improves by focusing on more number of hard positive samples. IEMOCAP $6$-way classification and MELD has are $6$ and $7$ class classification problems respectively and therefore the number of positives are not high for these datasets in a single mini-batch of size $32$. 
The improvement after adding  the sup-con loss is only $0.6\%$ and $0.3\%$ for IEMOCAP-6 and MELD (in absolute terms). When the number of classes reduce, the introduction of sup-con loss improves the performance of the model by a significant margin ($1.4\%$ and $1.1\%$ for IEMOCAP-4 way classification and CMU-MOSI respectively). 
This loss involves two hyper-parameters, namely the temperature and the weight used for combining with the cross-entropy loss.  We show the variation in the test performance with change in these parameters for IEMOCAP $4$-way classification and CMU-MOSI in Fig.\ref{fig:supcon}.

\subsection{Performance on different emotion classes}
In order to show the performance of our models in recognizing the different classes, we show the confusion matrices for the 6-way classification in IEMOCAP in Fig.\ref{fig:conf}. It is seen that the models have considerable error in differentiating happy and excited class. This is somewhat expected as these emotional classes are closely related to each other. The confusion matrices also highlight that the proposed model is not biased towards any particular emotion category class. 
\subsection{Role of test time ensembling}
\begin{figure}[t!]
    \centering
    \includegraphics[width=\columnwidth,trim={8cm 7cm 15cm 9cm},clip]{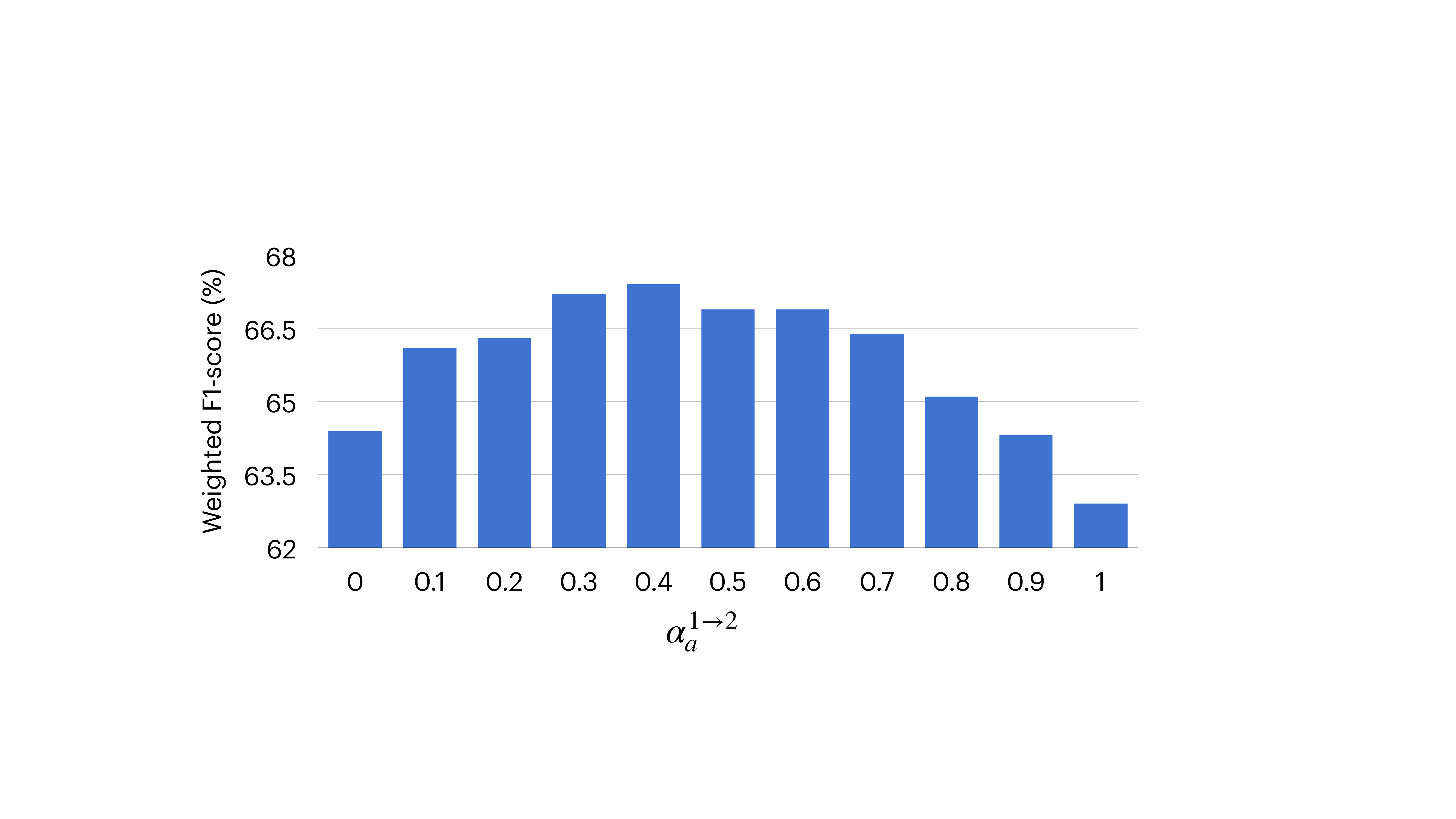}
    \caption{Variation of the test performance with change in $\alpha_a^{1\rightarrow2}$ (Eq.\ref{aud12}) for CMU-MOSI.}
    \label{fig:alpha}
\end{figure}

As mentioned in Sec.\ref{inference}, we take an weighted combination of the predictions of the model in a particular stage of training with those of the previous stage. The effect of this test time combination is shown in Table~\ref{ensemble}. We compare this strategy with the one when we do not use any ensembling in stage II and III of inference. We note that for CMU-MOSI, this improves the performance of the audio modality by $4.4\%$ in stage II. While a small improvement is noticed for MELD, the ensemble predictions does not yield any improvement for IEMOCAP. For the combination in stage III, we note that with the exception of CMU-MOSI, we see a slight improvement for all the other test settings. We show the variation of the test performance for the audio modality at the end of stage II for CMU-MOSI in Fig.\ref{fig:alpha}.
\begin{table}[t!]
\caption{\label{ensemble}Weighted F1-scores with and without inference time ensembling}
\begin{center}
\resizebox{\columnwidth}{!}{
\begin{tabular}{@{}c|c|cc|c|c@{}}
\toprule
\multirow{2}{*}{Modality} & \multirow{2}{*}{\begin{tabular}[c]{@{}c@{}}Inference Time\\ Ensemble\end{tabular}} & \multicolumn{2}{c|}{IEMOCAP} & \multirow{2}{*}{MELD} & \multirow{2}{*}{CMU-MOSI} \\ \cmidrule(lr){3-4}
 &  & \multicolumn{1}{c|}{4 way} & 6 way &  &  \\ \midrule
\multirow{2}{*}{Audio (Stage II)} & Yes & \multicolumn{1}{c|}{$78.7\%$} & $65.7\%$ & $50.1\%$ & $67.4\%$ \\ 
 & No & \multicolumn{1}{c|}{$78.7\%$} & $65.7\%$ & $49.6\%$ & $63\%$ \\ \midrule
\multirow{2}{*}{Text (Stage II)} & Yes & \multicolumn{1}{c|}{$81.4\%$} & $64.4\%$ & $65.6\%$ & $85.4\%$ \\ 
 & No & \multicolumn{1}{c|}{$81.4\%$} & $64.4\%$ & $65.4\%$ & $85.4\%$ \\ \midrule
\multirow{2}{*}{\begin{tabular}[c]{@{}c@{}}Audio + Text \\ (Stage III)\end{tabular}} & Yes & \multicolumn{1}{c|}{$85.9\%$} & $70.5\%$ & $65.8\%$ & $85.8\%$ \\ 
 & No & \multicolumn{1}{c|}{$85.3\%$} & $70.3\%$ & $65.5\%$ & $85.8\%$ \\ \bottomrule
\end{tabular}}
\end{center}
\end{table}
\section{Conclusion}\label{conclude}
In this paper, a hierarchical multi-modal neural architecture is proposed for conversational emotion recognition task. The proposed architecture improves the representation of the utterance level speech and text, first by learning the representations and then, by employing self-attention over other utterances in the recording. A co-attention mechanism is used for effective multi-modal fusion of the two modalities. On three benchmark datasets, we establish new state-of-the-art results. We further show the robustness of our model when tested with ASR generated text transcripts. Through extensive ablation studies, we also show the impact of different aspects of the modeling framework and the hyper-parameter choices. In future, we plan to extend these approaches to also incorporate the visual modality for emotion recognition. 
\bibliographystyle{IEEEbib}
\bibliography{refs}

% \newpage

 

% \end{IEEEbiography}



\vfill

\end{document}